\documentclass[useAMS,usenatbib]{mn2e}

\usepackage{multicol}

\usepackage{amsmath}    

\usepackage{graphicx}   

\usepackage{epsfig}

\usepackage{hyperref}
\usepackage{wrapfig}
\usepackage{url}

\title[STRESS - STEREO Transiting Exoplanet and Stellar Survey - I : Introduction and Data Pipeline]{STRESS - STEREO TRansiting Exoplanet and Stellar Survey - I : Introduction and Data Pipeline}

\author[Vinothini Sangaralingam and Ian R Stevens]{Vinothini Sangaralingam$^{1}$\thanks{E-mail: vs@star.sr.bham.ac.uk} and Ian R Stevens$^{1}$.\\
$^{1}$University of Birmingham, Birmingham, B152TT, UK.}

\begin{document}

\date{Accepted 2011. Received 2011 June; in original form 2011 January 04}

\pagerange{\pageref{firstpage}--\pageref{lastpage}} \pubyear{2011}

\maketitle

\label{firstpage}

\begin{abstract}

\indent The Solar TErrestrial RElations Observatory - \emph{STEREO}, is a system of two identical spacecraft in Heliocentric Earth orbit. We use the two Heliospheric Imagers (HI), which are wide angle imagers  with multi-baffle systems to do high precision stellar photometry in order to search for exoplanetary transits and understand stellar variables. The large cadence (40 min for HI-1 and 2 hrs for HI-2), high precision, wide magnitude range (\emph{R} mag - 4 to 12) and broad sky coverage (nearly 20 percent just for HI-1A and 60 per cent of the sky in the zodiacal region for all the instruments combined) of this instrument marks this in a space left largely devoid by other current projects. In this paper, we describe the semi-automated pipeline devised for the reduction of this data, some of the interesting characteristics of the data obtained, data analysis methods used along with some early results.\\
\end{abstract}


\section{Introduction}

\indent\indent Since 1572, when Tycho's nova was first observed, the astronomical community has come a long way in our search and understanding of stellar cycles and processes. Still at magnitude 12, it is estimated that at least 90 per cent of the variables are unknown \citep{Eyer}. Since \emph{Hipparcos} began the multi-epoch photometric survey, there have been many ground based  projects like the All-Sky Automated Survey \citep{Pojmanski}, Hungarian Automated Telescope (HATnet; \citealt{Bakos}), Wide Angle Search for Planets (WASP/SuperWASP; \citealt{Pollacco}) as well as the space missions like \emph{COROT}  \citep{Baglin}, Micro-variability and Oscillations of STars (\emph{MOST}; \citealt{Matthews}), \emph{Kepler}  \citep{Blomme}, Solar Mass Ejection Imager (\emph{SMEI}; \citealt{Spreckley}) which are dedicated to different types of variability studies and many of them have been hugely successful in not only discovering new exoplanets but also unravelling the stellar variability mysteries (\citealt{Michel, Pribulla, Basri}, etc.)\\  
\\
\indent In this paper, we present the use of twin Solar TErrestrial RElations Observatory (\emph{STEREO}) for the study of stellar variability and Exoplanetary transits. The STRESS project occupies a very interesting position in this scenario in terms of the location of its field of view, size of the field and magnitude coverage. The spacecraft covers a portion of the sky classically avoided for these type of studies, i.e., in the zodiacal path. As to the planetary transit searches, all of the dedicated space-based projects are concentrating away from this region. Hence this leaves a gap in the science for this particular sky region, which we are trying to fill in this project. Also, the pass band of this instrument lies very close to the \emph{R} magnitude of the stellar intensity spectrum. Hence we find that many of our tracked objects have actually no literature or very little basic information.\\
\\
\indent In Section 2, we describe the spacecraft and the instrumental characteristics, in Section 3, the data reduction pipeline and then the general characteristics of the data in Section 4 and some early result in Section 5 and summarising in Section 6.\\

\section{STEREO Spacecraft}
\indent\indent \emph{STEREO} is a set of two identical spacecrafts launched on October 25, 2006 to study the characteristics and evolution of Coronal Mass Ejections (CME) in stereoscopic projection. The two spacecrafts - one ahead of Earth (A) and one trailing behind Earth (B) - are in an Heliocentric orbit each separating from Earth at a rate of $22^{\circ}$ per year. A pair of Heliospheric Imagers (HI) in the SECCHI suite of instruments for each spacecraft is of particular interest to do stellar photometry due to its near continuous observation of the interplanetary region from 12 to 215 solar radii \citep{Howard}.\\

\begin{table}

\small

\begin{minipage}{80mm}

\caption{HI Instrument Characteristics}\label{tab1}

\begin{tabular}{|l||c|c}

\hline

 &\textbf{HI-1}& \textbf{HI-2}\\

\hline

Field of View& $20^{\circ}\times20^{\circ}$& $70^{\circ}\times70^{\circ}$\\

\hline

Plate scale & $35.15^{\prime\prime}$/pixel& $2.05^{\prime}/$pixel\\

\hline

Passband&6300\AA~- 7300\AA~&4000\AA~- 10000\AA \\

\hline



Nominal exposure time & 40 sec & 50 sec\\

\hline

Integrated cadence& 40 min & 2 hrs\\

\hline

\end{tabular}

\end{minipage}

\end{table}

\normalsize

\indent The HI-1 and HI-2 cameras have individual exposure times (equivalent to single shutter open/close cycle) of 40  and 50 seconds. Thirty such exposures are summed up  together on board the spacecraft, after correcting for cosmic ray effects, thus providing a cadence of 40 minutes for each HI-1 image transmitted to the ground. Similarly 99 exposures of HI-2 are summed to provide a digitally integrated image cadence of 120 minutes (see \citealt{Eyles} for detailed explanation). The actual $2048\times 2048 $ images are also $2\times2$ binned, resulting in $1024\times1024$ images for all the instruments, thus giving a final resolution of $70.30^{\prime\prime}$/pixel and $4.10^{\prime}/$pixel for HI-1 and HI-2 respectively.  \\

\section{Data Reduction}

\indent\indent {\bf{In this paper we describe the general analysis pipeline developed for all the HI instrument and specifically its application to the HI-1A instrument and the results derived thereafter. HI-1A instrument is preferred over the other HI instruments for this basic analysis due to its well defined PSF and also because it is devoid of many artifacts present in the other HI instruments. At a later stage, we plan to analyse and integrate the data from other HI instruments in order to increase the quality and quantity of light curves obtained.}} Here we describe the semi-automated photometry pipeline{\bf{ (see Fig.~\ref{fig:pipeline} for a schematic)}} developed with the help of the Rutherford Appleton Laboratory (RAL) using IDL solar soft packages using data for a period of 37 months from April 2007 till March 2010. Data from all the \emph{STEREO} spacecraft instruments are publically available for download from the UK Solar System Data Centre (UKSSDC) website\footnote{\url{www.ukssdc.rl.ac.uk}} in an interactive way allowing the user to download images for various periods of interest from the different SECCHI instruments.\\

\subsection{Pre-processing of data}

\indent\indent There are three different levels of processed data available - Raw summed images (L0), which are the on board summed images produced after removing the cosmic ray hits. The level one images (L1) are pre-processed using a pipeline developed by RAL \citep{Brown} to remove the various instrumental and general spacecraft effects explained below:\\

{\em{Flat-field}} - A flat field map, produced before the launch of the spacecraft during the ground calibration of instruments, is applied to the images to compensate for the differential gain across the CCD pixels.\\

{\em{Shutterless Readout}} - The HI cameras do not have a shutter, so that many short exposure images can be taken. A shutter less camera induces different exposure times to different parts of the image and smears it vertically. {\bf{A well understood mathematical correction of the form given below is applied to the image to decouple this effect from the images (See \citealt{Eyles} for a detailed explanation). The total response $R$ for a pixel $(n,m)$ is given by \\
\begin{eqnarray}
\lefteqn{R(n,m) = [T_{exp} \times I(n,m)] + \Sigma_{(y=m+1,2047)}[T_{clear} {}} \nonumber \\
         & & {}\indent \times I(n,y)] +\Sigma_{(y=0,m-1)}[T_{read} \times I(n,y)] \nonumber
\end{eqnarray}
\\
Where  the count rate for the pixel $(x,y)$ is given by $I(x,y)$, $T_{exp}$ is the exposure time, $T_{clear}$ and $T_{read}$ are the line transfer time during the previous exposure clearing and current exposure reading.}}\\

\begin{figure*}
\centering
\begin{tabular}{c c}
\includegraphics[scale=0.65]{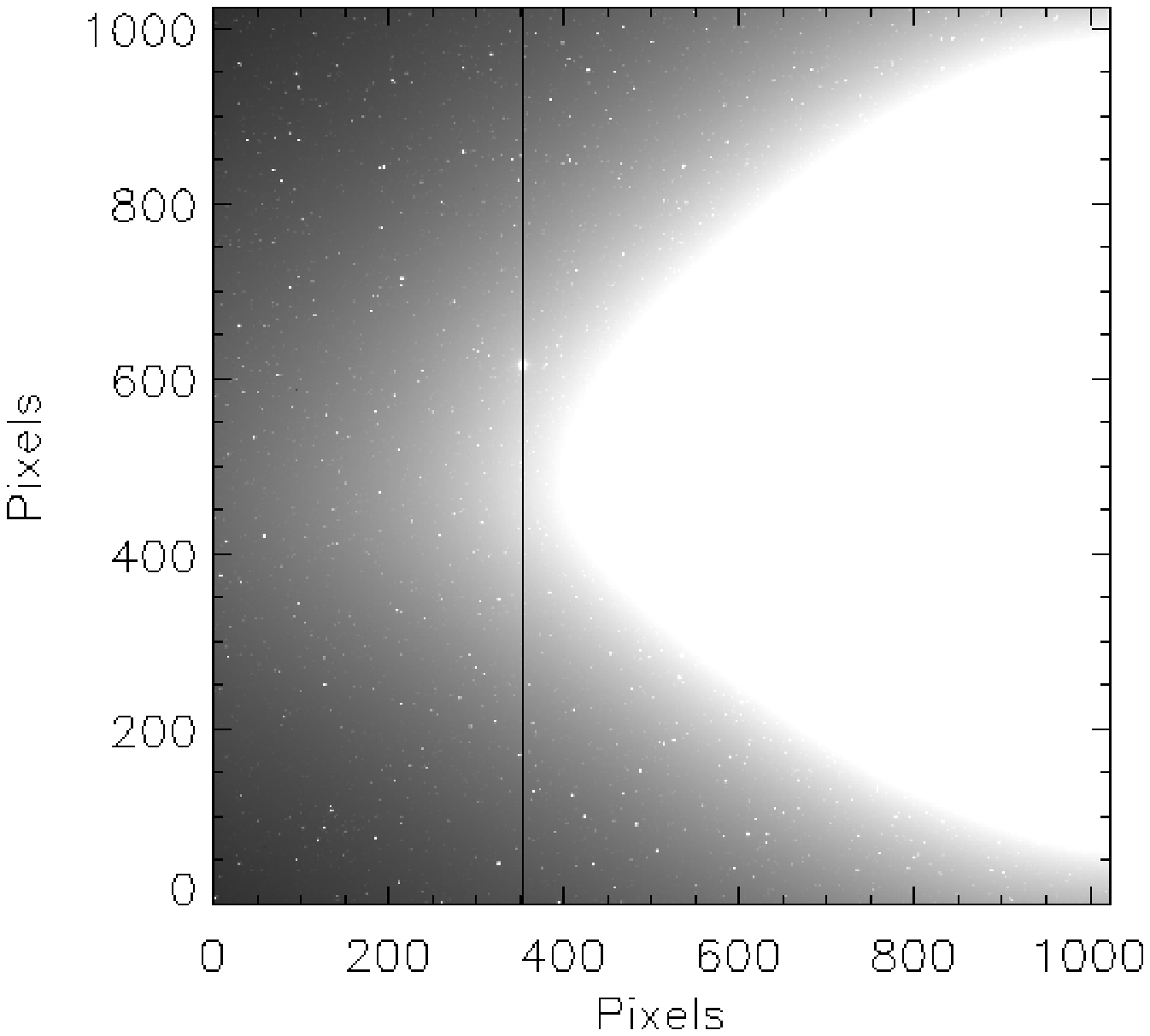}&\includegraphics[scale=0.65]{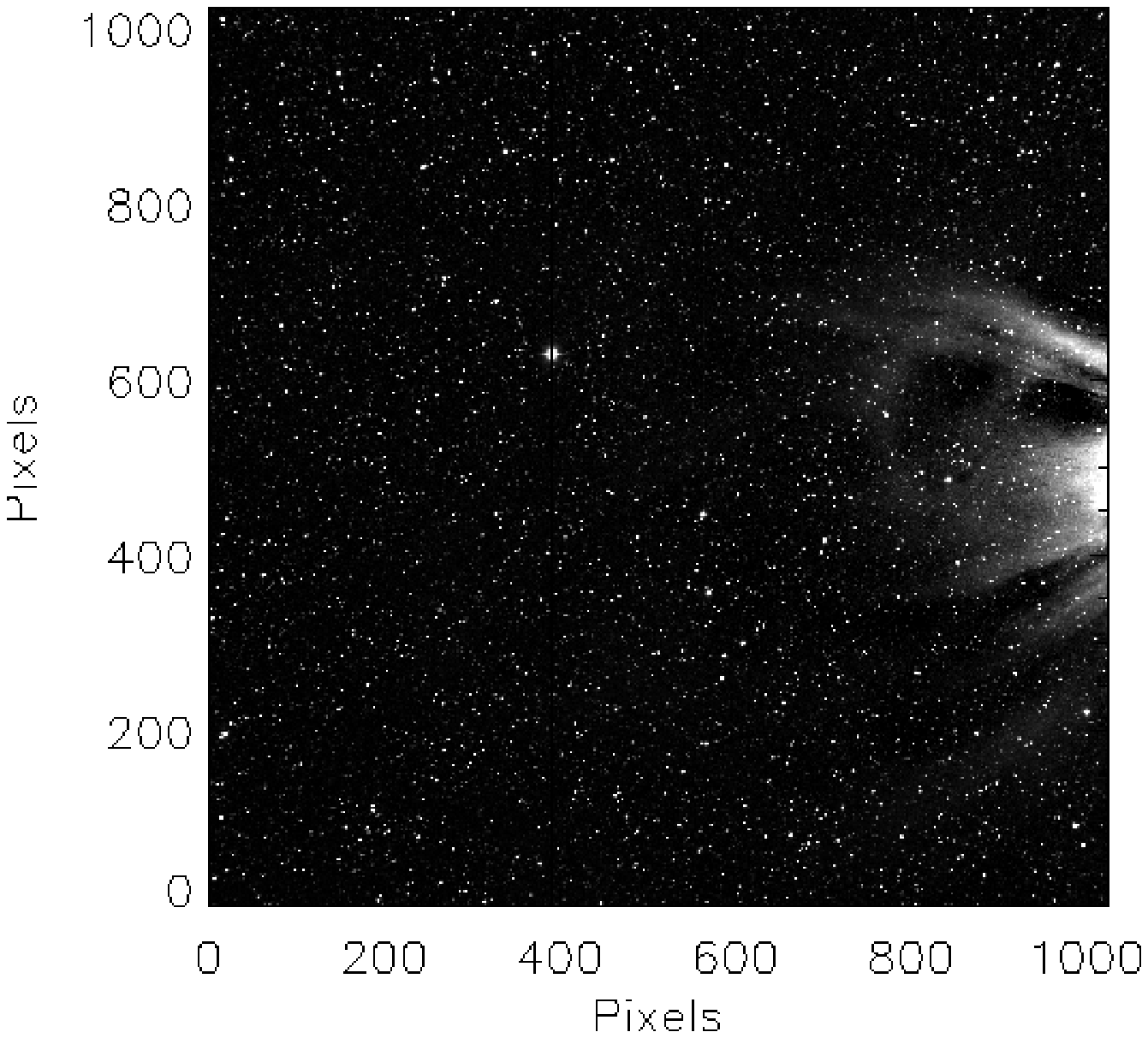}
\end{tabular}
\caption[Illustration of the dominance and need for F-coronal subtraction.]{\label{fig:with_without_corona}Left - Sample image with the solar F-corona present. The line running across the image is the blooming effect caused by a planet which can be seen prominently in the image on the right. Right - The same image after the subtraction of a running window map of one day which removes most of the contribution from the F-corona, thus enabling us to do stellar photometry. We can also identify the planet as well the solar streamers in this image, which were impossible in the image on the left. The image in the right is the data image that we use in further analysis. We can also understand from this figure the influence of the solar phenomenon in the photometric reduction and analysis.}

\end{figure*} 

{\em{Missing Data}} - Blocks of data which were lost due to telemetry drop-out are replaced by NaN values in the images.\\

{\em{Bad Files}} - Certain images are tagged as bad due to various reasons like corrupted data, telemetry dropouts, split star field, integer overflow, etc. These are identified using a bad file list provided by the UKSSDC and removed from further analysis.\\

{\em{Blooming}} - For magnitude brighter than 4 ($m<4$), HI-1 pixels saturate and HI-2 pixels saturate for magnitudes brighter than 1 ($m<1$). Such columns are identified and replaced with NaN values.\\

{\em{Updated Pointing Information and Optics Parameters}} - Raw data contains only nominal pointing information and camera optics as calculated from pre-launch parameters. In the initial stages of analysis, the instrument pointing values were found to have larger offsets than those predicted. Hence an updated spacecraft pointing, along with optical parameters, is calculated using star fields in the image and the header is updated with the new information during this stage.\\

\indent Even after this process, we found that certain frames cannot be used for further reduction due to various reasons. The most common cause among them was found to be due to certain spikes and black outs in the data which are not tagged by the previous routines and these bad frames were removed automatically using a five sigma clipping.\\

\subsection{F-corona subtraction}

\begin{figure}
\centering
\includegraphics[scale=0.5]{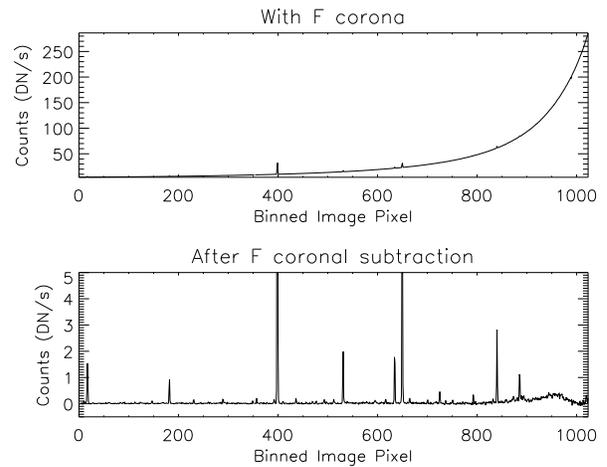}
\caption[Image slice]{This is a slice of the image through the centre before and after the F-corona subtraction. The spikes are the bright stars in the field which are barely visible in the top image and we can identify even the fainter stars when the F-corona is removed.} 
\label{fig:image_slice}
\end{figure}

\indent\indent The data at this stage (L1) still has the dominant solar F-corona, which is the scattered sunlight from the dust particles in the inner Heliosphere. This light extends till Earth's orbit and merges with the zodiacal light at night. The brightness of this corona varies from $10^{-12}$ to $10^{-6}$ times the brightness of the solar disk. Fig.~\ref{fig:with_without_corona} illustrates the need for the removal of the F-corona in order to do stellar photometry. The dark line running through the column is due to blooming of objects brighter than $m = 3$, which can be seen prominently in the figure on the right. Fig.~\ref{fig:image_slice} contains a plot of the image through the middle row with the F-corona present and when it is removed. It is clear from this plot the dominance of F-corona in the data. The stars which are almost invisible in the top plot becomes clearly visible in the bottom plot. The brightest of the objects at around 400 and 600 pixels are 6th magnitude stars and the fainter ones which are barely visible in the bottom plot are the 11th and 12th magnitude stars. During the initial stages of the project, a minimum background map is obtained by computing the minimum value of each pixel for a set of images. After experimenting with different number of images {\bf{and studying the residual F-corona and its variation present in the data, we found a day worth of images, i.e., 36 images, to be a best fit.}} This map is then subtracted from each of the images which removes most of the F-corona. \\

\begin{figure}

\centering
\includegraphics[scale=0.45]{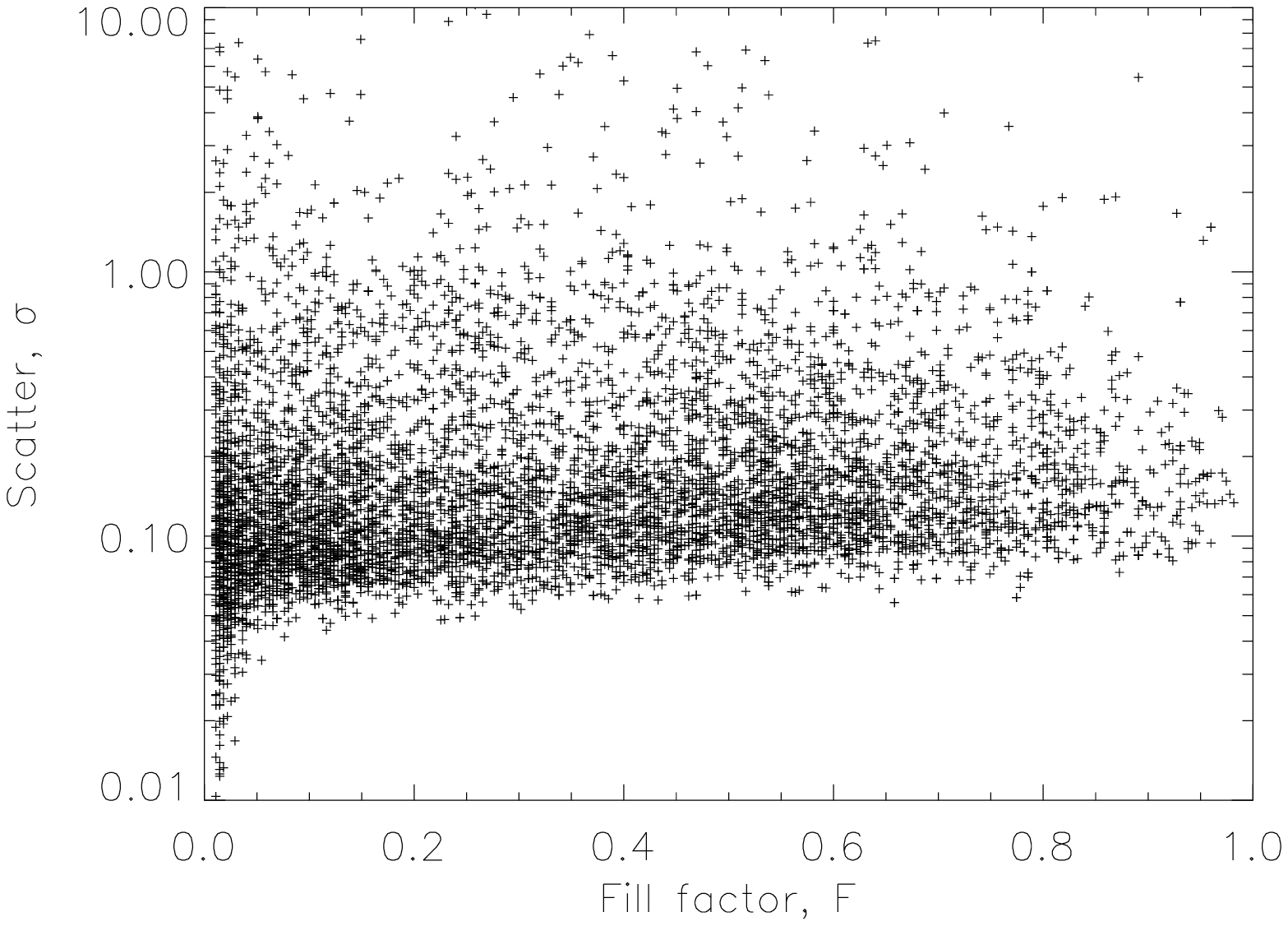}

\includegraphics[scale=0.45]{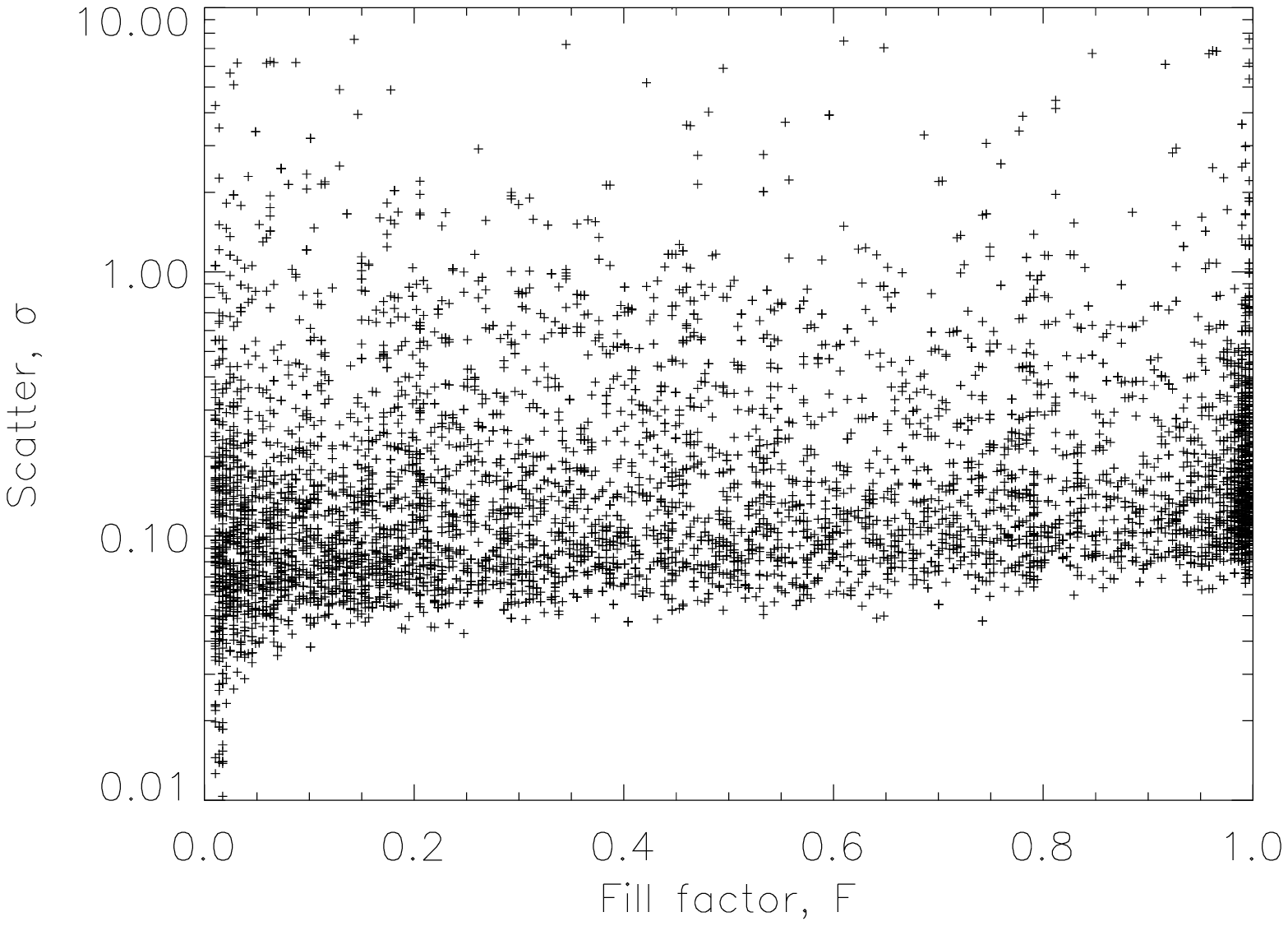}

\caption[Methods of F-corona removal]{Plots of scatter vs fill factor for a sample data of around 7700 stars tracked for 10 days in the month of October 2007. The top plot is using the one day minimum map as a background and the bottom plot is using the new running window as the background for F-corona subtraction (see text for further explanation).} 
\label{fig:old_new_bg}

\end{figure}

\indent The light curves computed after this method of F-coronal removal has a dominant signal of around one day corresponding to the minimum map used for subtraction and this made studying periods around one day impossible. Recently, RAL\footnote{\url{http://www.ukssdc.rl.ac.uk/solar/stereo/documentation/HI_processing_L2_data.html}} formulated a new method of F-corona subtraction using a running window either of one day or eleven days and within that window select the lowest 25 per cent of the data and form an average of this data on a pixel by pixel basis, which is then subtracted from the images. Two different window lengths are used in order to study different types of solar phenomenon. In the case of stellar photometry, the one day window seems to improve the photometry of the data much better than the eleven day window. To further understand the effect of this new background subtraction in our data, we computed a parameter, the Fill factor ($ F $), {\bf{which essentially gives the percentage of data points available for an individual object or how well the star is tracked through a sequence of images. This is computed by dividing the number of data points present in the final light curve for the star by the total number of frames present from the moment when the star enters the field of view to when it leaves.}} In Fig.~\ref{fig:old_new_bg}, we plot this fill factor, $ F $ against the standard deviation, $\sigma$ of the light curves.\\

\indent The data used in this plot is from around 7700 stars which are tracked when the instrument faces the galactic centre and hence involves a very crowded field, in order to consider an extreme case. The $\sigma$ are computed from raw light curves which contain the long term trends mainly caused by the variation of F-corona in time scales of the data. {\bf{A fill factor of one indicates that the star is completely tracked through all the images and $F < 1$ indicates a loss of data during its presence in the field of view. We find that in the running window method, there are many stars which are completely tracked whereas in the old background subtraction method none of the stars are tracked completely.}} Also from the data, we can infer that the 'raw' scatter of the data is close to 10 per cent photometric precision. There are a few objects below that scatter, close to the zero value of the fill factor, which are the result of very few data points available for computation and are not indicative of better photometry. But it has to be pointed out that this precision is obtained with all the systematic effects present in the data and hence there is a huge scope for improvement after treatment with various trend-removing methods. We did seem to reach a $\sigma$ of less than 1 percent when the systematics are removed as discussed in Section \ref{data_char_sec}.\\

\subsection{Stellar photometry}\label{pipeline_section}

\indent\indent Photometric information is now extracted from these images (known as L2 data) using the NOMAD \citep{Zacharias} as input catalogue for stars {\bf{brighter than $R = 12$ by our semi-automated pipeline which is explained below}}. \mbox{NOMAD\footnote{\url{www.nofs.navy.mil/data/FchPix/}}}\@ -  Naval Observatory Merged Astrometric Dataset has information for nearly 1.1 billion stars derived from Hipparcos, Tycho-2, UCAC-2 and USNO-B1 catalogues for astrometric and optical data and 2MASS catalogue for infra-red photometric information. Some of the catalogues are complete up to a \emph{V} magnitude of 21 \citep{Monet}.\\ 

\begin{figure}

\centering

\includegraphics[scale=0.2]{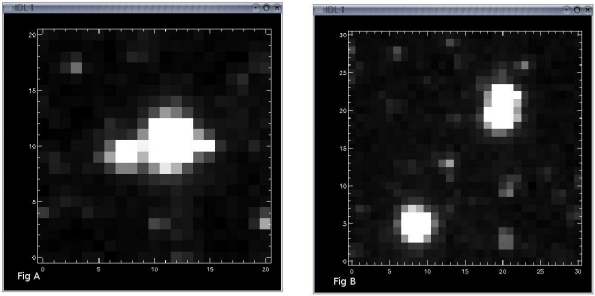}

\caption[An example of PSF contamination in a crowded region]{\emph{STEREO} HI-1A images of stars. Left panel: An example of two stars which are at a distance of less than 10 pixels contaminating each other and have a magnitude difference of 2.5 or lesser. Right panel: An example of two stars of comparable magnitude which are well separated such that their PSF does not overlap and are well distinguished at a separation of 10 pixels.}

\label{fig:stars}

\end{figure}

\indent We use the image header information to convert from the spacecraft to sky co-ordinates and thus identify targets using our input catalogue. The pixels are flagged as probable stellar sources, if the count is 1.5 times higher than the surrounding 4 by 4 region. When these sources are matched to the catalogue we use an error margin of 2 pixel.{\bf{ The large pixels of the HI images induces PSF contamination between close targets of comparable magnitudes. We apply a basic magnitude-distance criterion to identify and remove a large portion of this effect. Thus, pair of objects which are closer than 10 pixels and have a magnitude difference of less than 0.5 are removed from the target list. We also remove the fainter star in these pairs within the same radius but with a magnitude difference of 2.5 or lesser. This is explained in Fig.~\ref{fig:stars}, where in the left panel, there are two stars which are less than 10 pixels away and are of comparative brightness that they contaminate each other and in the right panel, there are two stars which are clearly distinguishable when separated by more than 10 pixels. In the case of bright stars, this just resolves the two PSF. In order to detect further contamination, which is not removed by this filter, we use three different aperture sizes for calculating the flux. By using the biggest aperture for the brighter star and the smallest for the fainter star, we try to rule out further confusion. In spite of all these measures, we still find that targets are either missed out or mistaken for nearby closer and brighter ones. We discuss these issues in detail in Section {\ref{sec:post_process}}. We would also like to note that in all these process, we are only dealing with the correct identification of targets and the issue of dilution of flux by nearby targets along with a noise analysis will be discussed in detail in a forthcoming paper.}}\\

\indent Generally, the PSF of HI-1A is well behaved in most regions of the CCD, although it deviates a little towards the edges of the frame. The PSF in the centre of the field is highly circular whereas it becomes elliptical at the edges. This effect is more prominent in other HI instruments. Hence, a more robust method of identifying the centroid of the targets is essential., Therefore, we use the method developed by Steve Spreckley for use with \emph{SMEI}, where the PSF is of a very odd shape \citep{Spreckley}. Aperture photometry is done on these selected targets with the inbuilt IDL routine for aperture radii  2.5, 3.0 and 3.5 pixels and sky background is calculated between pixel radii 6.0 and 10. The CCD is assumed to have a quantum efficiency of 90 per cent for this spectral range and the photon-electron conversion gain {\bf{for the camera}} is 15 units  \citep{Bewsher}. Sky variance, random photon noise and the uncertainty in the mean sky brightness contribute to the measured error values of the flux. {\bf{The flux values mentioned in this paper are in units of $DN/s$ and not converted to a magnitude scale.}}\\

\begin{figure*}

\centering
\includegraphics[scale=0.5]{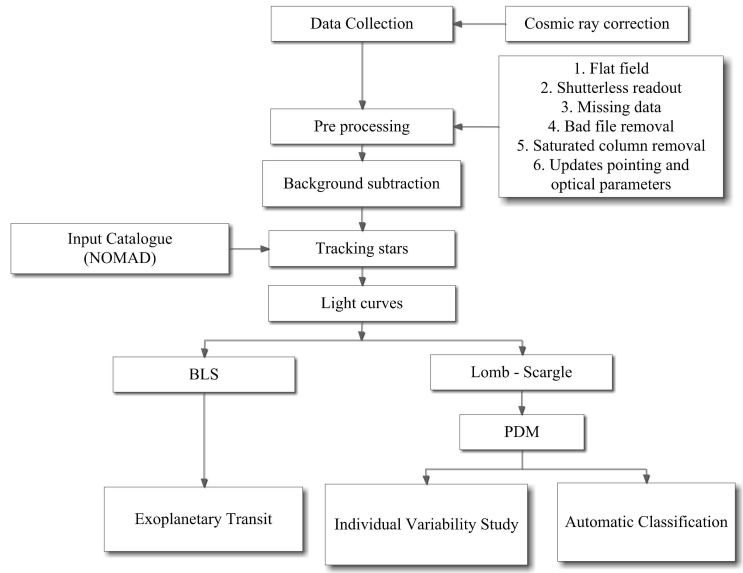}
\caption[STEREO HI data analysis pipeline]{A schematic of the STEREO HI data analysis pipeline, showing the various stages of data preparation involved in producing light curves from raw images (see text for explanation). }
\label{fig:pipeline}
\end{figure*}

\indent The time stamp in the final data is the time at the start of the observation in Geocentric Julian dates with correction for the sun-spacecraft distance applied. This light travel distance between the sun centre and the spacecraft is found from the image header keyword. We would like to note that this time of observation will vary across the CCD due to the large sky area covered but this difference would be very small compared to actual cadence and hence ignored. Also, due to the complicated integration of the data on board the spacecraft, the start of observation is found to be the best parametric representation of time. Fig.~\ref{fig:pipeline} gives a schematic of the analysis pipeline discussed so far.\\

\indent In the next section, some of the general characteristics of the data are explained in detail.\\

\section{Data Characteristics}\label{data_char_sec}

\indent\indent Complete data for each star, which comprises of the Right Ascension (RA) and Declination (Dec) marking the position, \emph{R} magnitude obtained from the input catalogue, flux values computed from the pipeline for each aperture and its error values along with the sky background for each frame and its error values, is stored in an IDL data format. Typical time taken by an object to transit the HI-1 field of view is about 20 days (around 700 data points) and for HI-2 instrument it is about 70 days (around 2200 points). An integration of data from all instruments will give us around 180 days worth complete data set for each orbit of the spacecraft. Since the cadence of the {\bf{HI-1}} instrument is 40 minutes, the Nyquist frequency for this set of data is around 18 cycles per day (c$d^{-1}$) and {\bf{for HI-2 it is 6 c$d^{-1}$}}. So far we have obtained data for nearly a million stars covering a period of 3.5 years.\\

\indent The data in general exhibits systematic trends which are the resultant of the basic instrumental design as well as trends which are common to all the stars in the region either due to some part of the processing pipeline or due to a systematic change in the basic conditions of the images.\\ 

\begin{figure}

\centering

\includegraphics[scale=0.5]{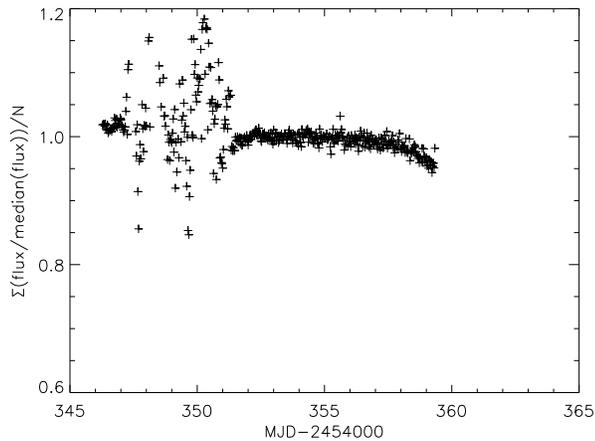}

\caption[Illustration of the effect of Solar events on light curves]{A plot of the median value of all the light curves (12) against time in a small strip of 10 pixels by 1024, which encounters a Solar wind transient event. This clearly illustrates the effect of such events in all the objects in the region.}
\label{fig:cme_noise_lc}
\end{figure}

\indent One of the most common phenomenon is the increase or decrease in the median intensity of the flux due to the presence of either a Solar-wind transient event or a CME in the image. Since the flux value is relative to both the actual pixel intensity as well as the different backgrounds which are being subtracted in various stages, the increase in intensity can be attributed to an increase in the flux of the whole region and decrease to a higher value of coronal background which is being subtracted and hence we find this effect being manifested in all the light curves in the region. In order to illustrate this effect, we have plotted the median intensity of all the light curves in a particular region for a period in which we know the camera encountered a Solar transient event \footnote{The event list is provided at the STEREO site \url{http://www.sstd.rl.ac.uk/stereo/HIEventList.html}} in Fig.~\ref{fig:cme_noise_lc}. The data used in this plot is from a very small strip-like region of CCD comprising of 10 pixels by 1024 pixels and we can clearly identify the time when the Solar wind cloud passed through this region imaged by the CCD from the high scatter of all the light curves in that portion. We calculate the median by first normalising the light curves with its median value before adding all the intensities and dividing by the total number of light curves.\\

\indent Fig.~\ref{fig:cme_noise_lc} also shows another common trend, which is the loss of flux at the end of the frame. Sometimes such a loss is also found at the start of the light curve. These effects are due to the variation of the PSF at the end of the CCD and are magnified in the other HI instruments. Another trend we commonly find is a systematic variation across the whole light curve for all the light curves. Since we can remove this trend by fitting a 4th order polynomial, which is the same as the CCD response function, we think this effect is a manifestation of the response function \citep{Bewsher}. A detailed discussion of this trend can be found in the next section.\\
\\
\indent Another feature which we find in many light curves are sudden flux increase, similar to flares - a sharp and sudden increase for a few frames, mostly caused by the passage of the object either through a solar filament or through a bright object like a planet or a brighter star. These are highly transient and temperamental.\\
\\
\indent All these above trends, though systematic, are highly time dependent. But there are also trends which are independent of the time of the observation but rather related to the magnitude of the object. This is the relative scatter of the flux against magnitude of the star. In an ideal case, when we remove all the systematic noise mentioned above, we can end up with the photon-shot noise limited flux values corresponding to the magnitude of the object under study and one of the advantages of a space based mission is that we can easily identify all the systematics and hence can achieve this limit much more easily than a ground based observation. But in this project, we are limited at a higher flux value due to the basic instrumental aim of studying solar phenomenon and not stellar variability analysis. Some of the limiting factors in this case are the presence of bright solar corona and planets, large image pixel, Galactic Plane crossing and last but not the least, a very small team dedicated to its analysis. \\ 

\begin{figure}
\centering
\includegraphics[scale=0.5]{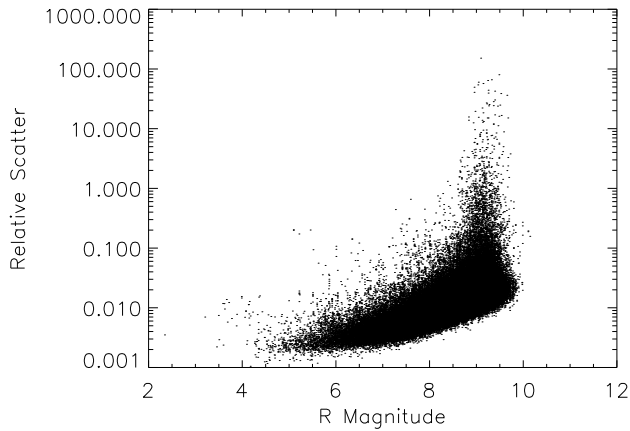}
\caption[Plot of relative scatter against magnitude]{A plot of relative scatter of intensity against magnitude of the stars.}
\label{fig:sig_mag}
\end{figure}

\indent Fig.~\ref{fig:sig_mag} is a plot of the relative scatter (by relative scatter we mean the variance found after normalising the light curves) against \emph{R} magnitude of stars tracked during September and October of 2008. The sample is representative of general STRESS characteristics since the later part of the sample is through the highly crowded galactic plane region and the former is through a relatively sparsely distributed region. A one day boxcar average is smoothed from all the light-curves before calculating the scatter. This is to remove the residual effects of the F-coronal subtraction. Hence most of the variability in the data is either stellar or other systematics. Still, for most of the objects brighter than 8th magnitude, we have a relative scatter less than 1 per cent and for fainter objects a few percent.\\

\subsection{\label{sec:post_process}Post processing the Light curves}

\indent\indent From the above section, we find that there are many systematic variations present in the light curves. In this section, we present some of the methods attempted and finally used in removing or reducing these effects so as to analyse for periodicity in these data. Some of the well established methods in use for this type of analysis are the filtering methods like median smoothing and box-car average smoothing, polynomial fitting and various new specialised algorithms like Sys-Rem \citep{Tamuz}, Trend Filtering Algorithm (TFA; \citealt{Kovacsb}, Dae Kim-won's algorithm \citep{Kim}. Even though these form the basis of many photometric analysis, they have to be adapted specifically to our light curves due to the scanning nature of our instrument and the different background issues in the data. One of the main issues we have to deal with in our data, is the crowded field. Even after application of the confusion criteria mentioned in Section \ref{pipeline_section}, we find that we are tracking two objects alternatively and this raises the scatter of the light curves. Usually this confusion occurs when we have two or more nearly similar brightness objects. Since we are applying the criteria for only pairs of objects, sometimes these objects are still tracked.\\

\begin{figure}
\centering
\includegraphics[scale=0.52]{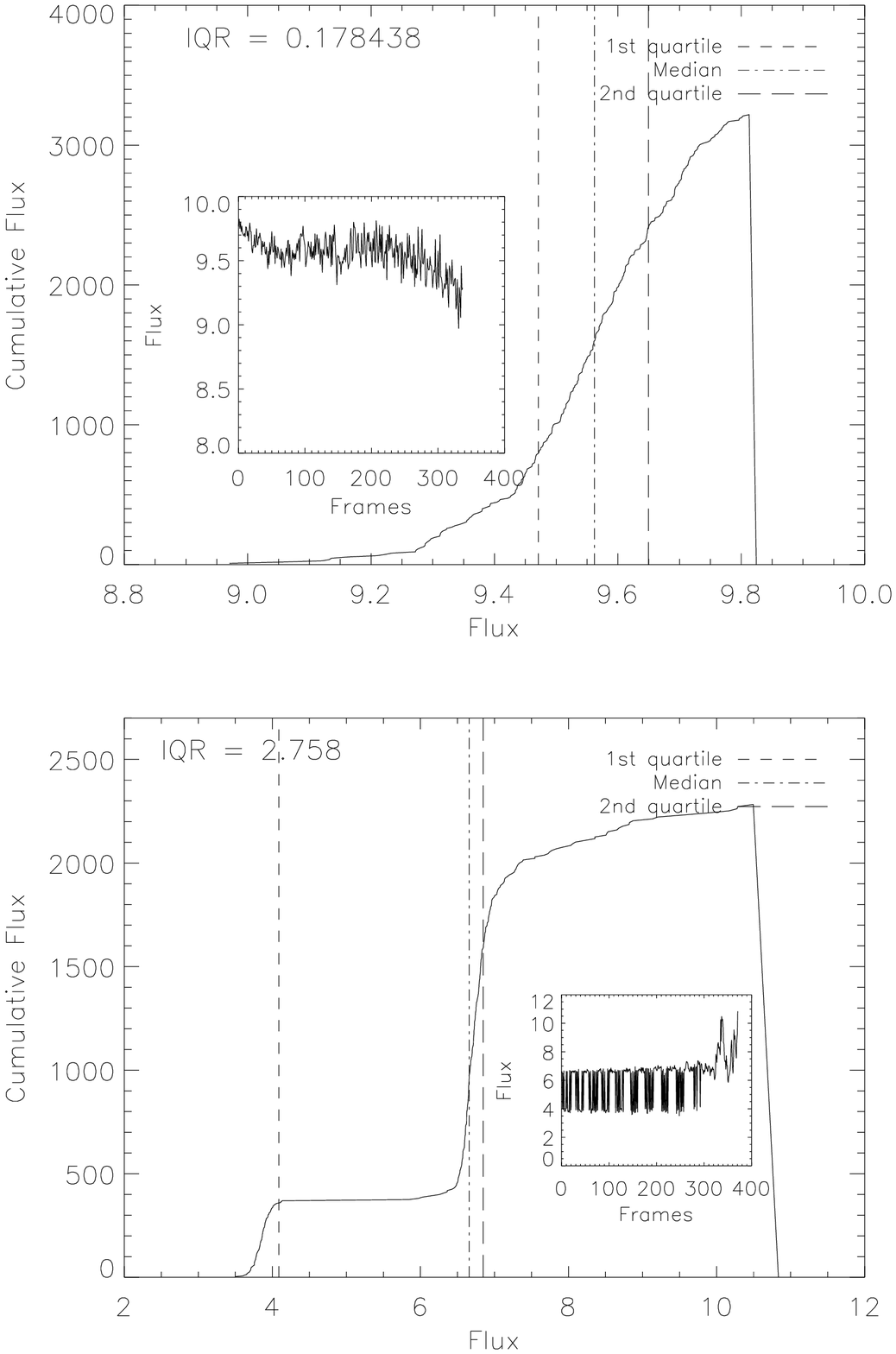}
\caption{Quality control in light curves: Above, we have plotted the cumulative frequency distribution of a light curve against actual intensity. The top plot is the distribution for a 'good' light curve, i.e, where only a single star is tracked throughout the CCD. The actual light curve is given as an insert in the plot. The bottom plot is the same distribution when two stars are tracked alternatively such as to give rise to a periodic signal. We can easily identify the 'bad' light curve from the inter-quartile range (IQR) which is given. The higher this value, the larger the spread of the light curve. This is a fairly robust method of identifying wrongly tracked objects in our data.}
\label{fig:IQR}
\end{figure} 

\indent Even though these light curves make up less than 1 per cent of the total, these 'alternating' light curves generally create a very periodic signal, since the tracking algorithm shifts the centroid very systematically depending on which star is in maximum at that point and hence often these light curves get flagged up as a periodic signal during the period analysis. Even a robust sigma clipping does not identify these objects since the actual variance of the light curve itself is very high and thus is an inaccurate measurement of the scatter in the system. In order to flag these objects, we use the inter-quartile range measurements. Fig.~\ref{fig:IQR} illustrates this method in detail. The top plot is the cumulative intensity against flux for a good light curve (see insert) and the bottom is the same for a highly scattered light curve, which is the tracking of two objects almost alternatively. Another method which was used initially as a cross check, is the difference in intensity for the various apertures. The fainter stars have the same flux value irrespective of the aperture size and the brighter stars just increase their flux values as we go to larger apertures since most of the star PSF is captured in larger apertures. But when there is a confusion between two stars, there is a random change in flux values irrespective of the increase in aperture size. Hence, they can be used to identify wrongly tracked objects.\\

\begin{figure}
\centering
\includegraphics[scale=0.57]{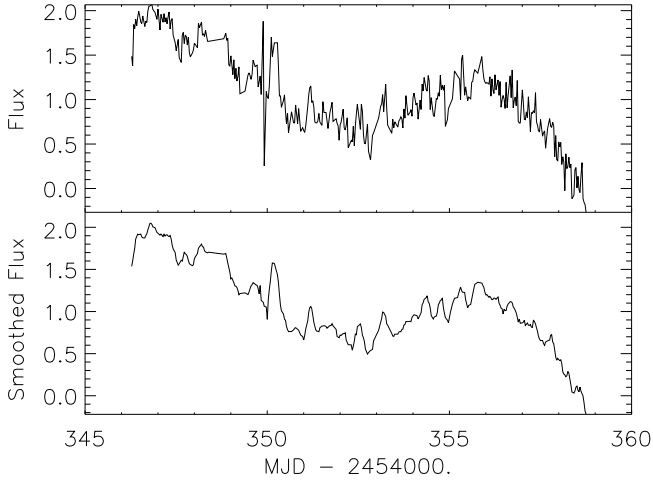}
\caption{The top plot is the original light curve of \emph{V} = 6.7 star before applying any correction and the bottom plot is the 3 hr smoothed function which is used as the light curve for further analysis. This ensures the removal of the high frequency noise in the data. This filter is applied only for stellar variability analysis and not for planetary transit analysis.}
\label{fig:highfreqnoise}
\end{figure}

\begin{figure}
\centering
\includegraphics[scale=0.5]{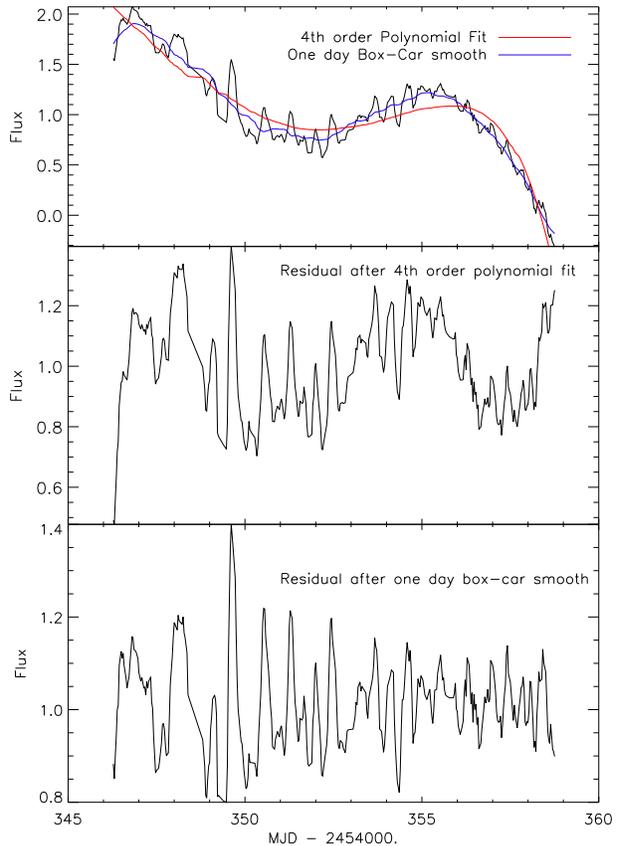}
\caption{The top plot illustrates how both a 4th order polynomial as well as a one day box-car average smoothing function fits to the light curve. The bottom two plots give the residuals after removing the polynomial and the box-car filter respectively.}
\label{fig:smoothing}
\end{figure}

\indent The light curves selected after using the quartile filter, has both a high frequency and a low frequency noise component in them. We first remove the high frequency noise by using a box-car filter corresponding to a box width of around 3 hrs. Since this is also the time scale of any planetary transiting event, we implement this filter only for stellar variability analysis. We use the resultant smoothed function for further analysis and the residual in the calculation of our signal-noise ratio when a stellar signal is confirmed. Fig.~\ref{fig:highfreqnoise} illustrates how the use of smoothed function removes the high frequency noise in the data for a randomly selected standard star (HD 112281). This has a \emph{V} mag of 6.7 \footnote{All the magnitude mentioned hereafter refer to information from the Simbad database for ease in cross checking.} and a fairly non-interesting star, which will be used in all further processing examples. The usage of this smoothed function as the light curve for further analysis does not significantly affect the standard deviation of the light curve.\\

\indent In order to remove the long term trends, two different methods were experimented with - a box-car average smoothing and polynomial fitting. After trying different box widths, a one day smooth was found as a best fit to many light curves. Removing a 4th order polynomial also tend to remove most of the long term trend in the data. From \citet{Bewsher}, we know that the CCD response varies exactly as a fourth order polynomial across the CCD central axis. And hence could be causing this systematic trend across all the light curves. One of the disadvantages of the polynomial fitting is that, when there is a larger stellar variability signal, this fitting becomes inaccurate and thus increases the variance of the final light curve. Even though, the polynomial is a theoretical fit to the trend, when the pipeline is automated, box-car smoothing is more reliable so as to remove only the systematics. Fig.~\ref{fig:smoothing} shows how a box-car average of one day and a 4th order polynomial fits the long term trend in this sample light curve. We also find that the polynomial fit does not remove all the long term trend but only that specified by the CCD function whereas the smooth function removes all types of long term variation. This is certainly helpful when looking for periods relative to the length of the light curve.\\

\begin{figure*}
\centering
\begin{tabular}{c c}
\includegraphics[scale=0.5]{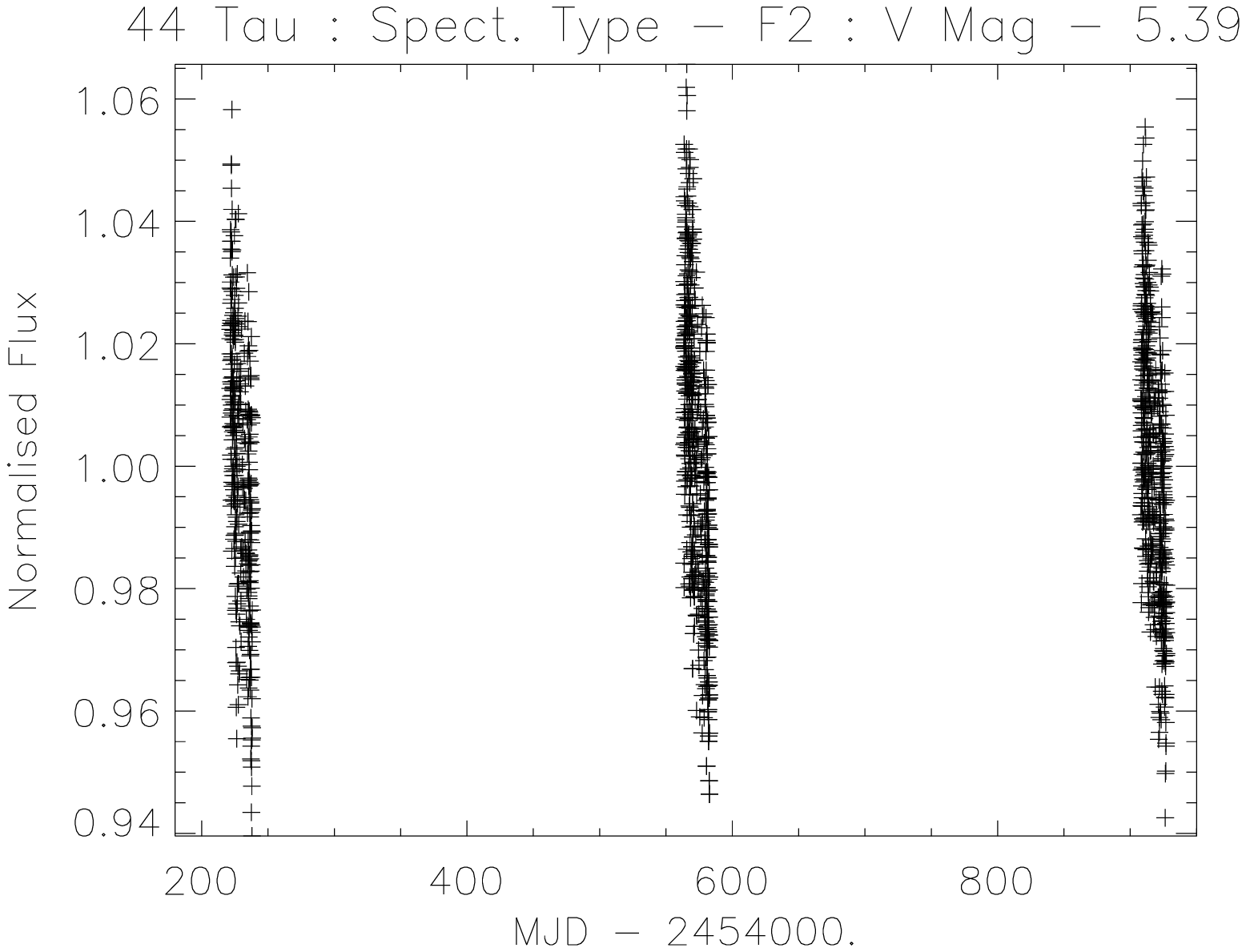}&\includegraphics[scale=0.5]{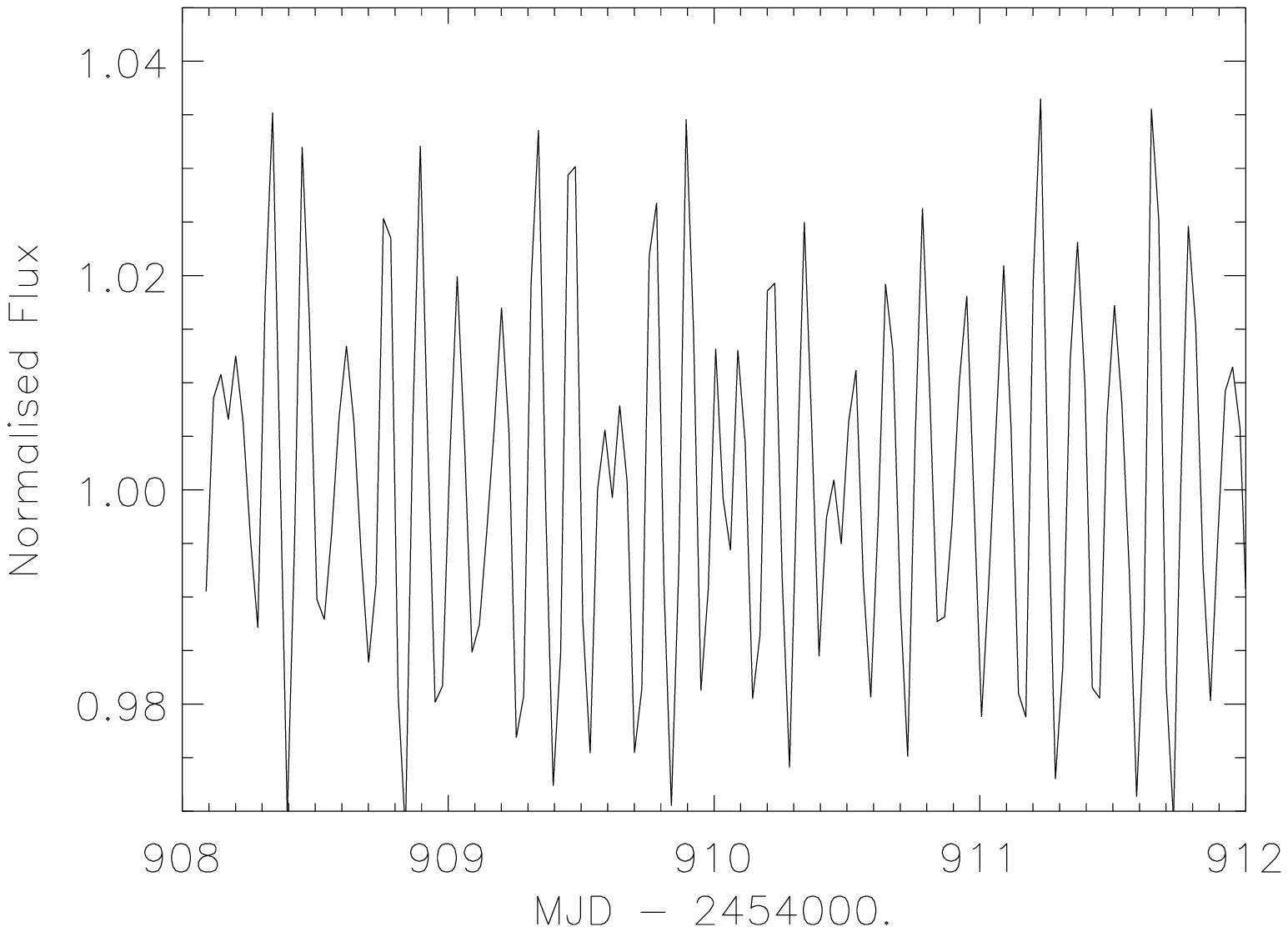}
\end{tabular}
\caption{\label{fig:44_tau} The above is the STRESS light curve of the 44 Tau system. The plot on the left is an illustration of the data availability for more than three epochs as of now. This aids in more precise period determination as well as characterising period changes in an object. The right plot is a zoomed in version displaying clearly its symphony of frequencies. } 
\end{figure*}

\indent After removing the various known noise sources in the light curve, we analyse them for variability using Lomb-Scargle periodogram (LS; \citealt{Scargle}) and Phase Dispersion Minimisation (PDM; \citealt{Stellingwerf}). We use LS to flag up potential variables and then confirm the periodicity with PDM. Our initial analysis found that the LS is very sensitive to larger periods while PDM is sensitive to the shorter periods and thus using both increased the probability of finding a true variable and eliminating false positives. We also used  Box-fitting Least Squares method (BLS; \citealt{Kovacsa}) to look for exoplanetary transits. All of these periodogram analysis will be described in detail in a forthcoming paper. \\
          
\begin{figure*}
\centering
\begin{tabular}{c c}
\includegraphics[scale=0.5]{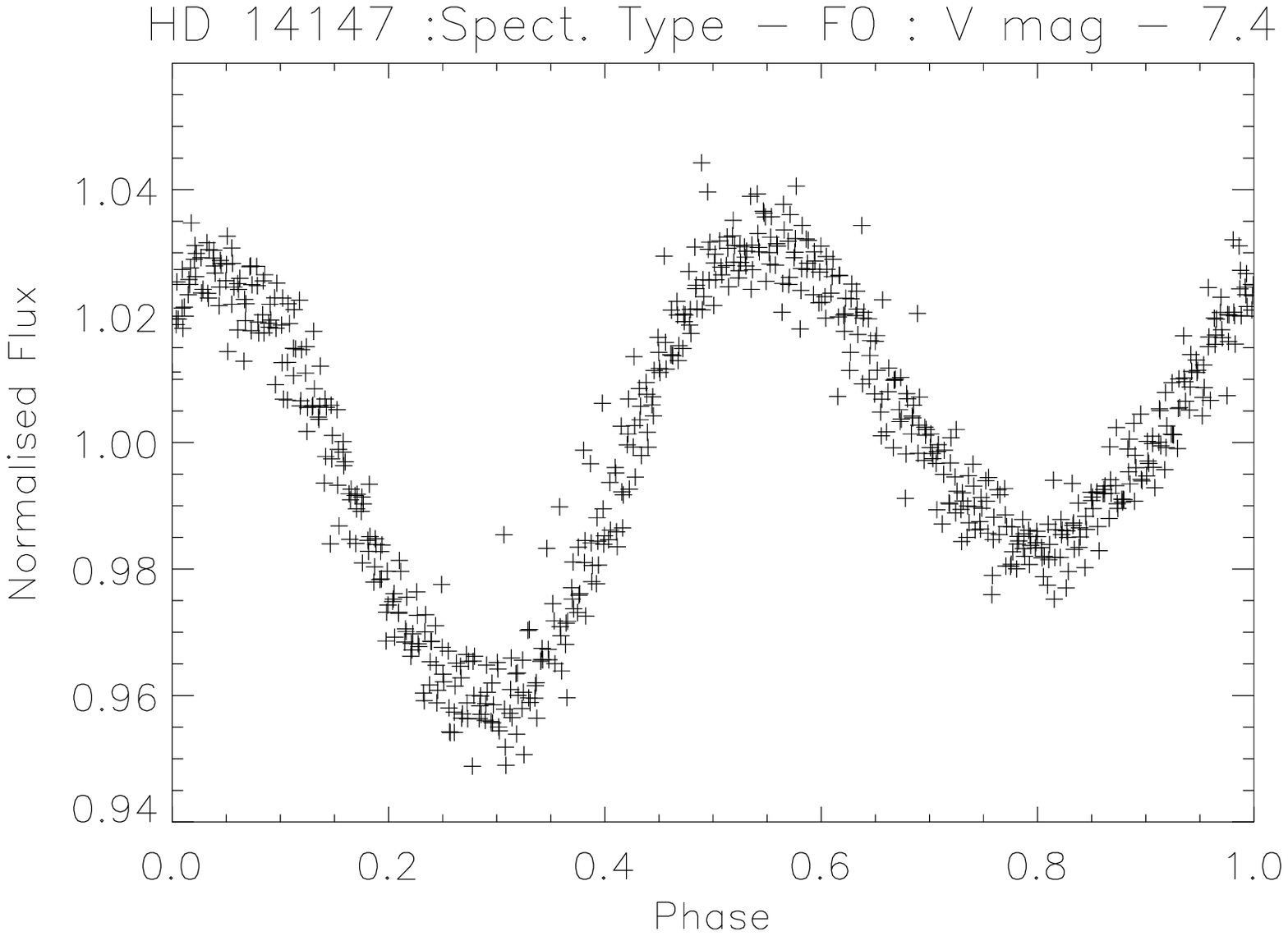} & \includegraphics[scale=0.5]{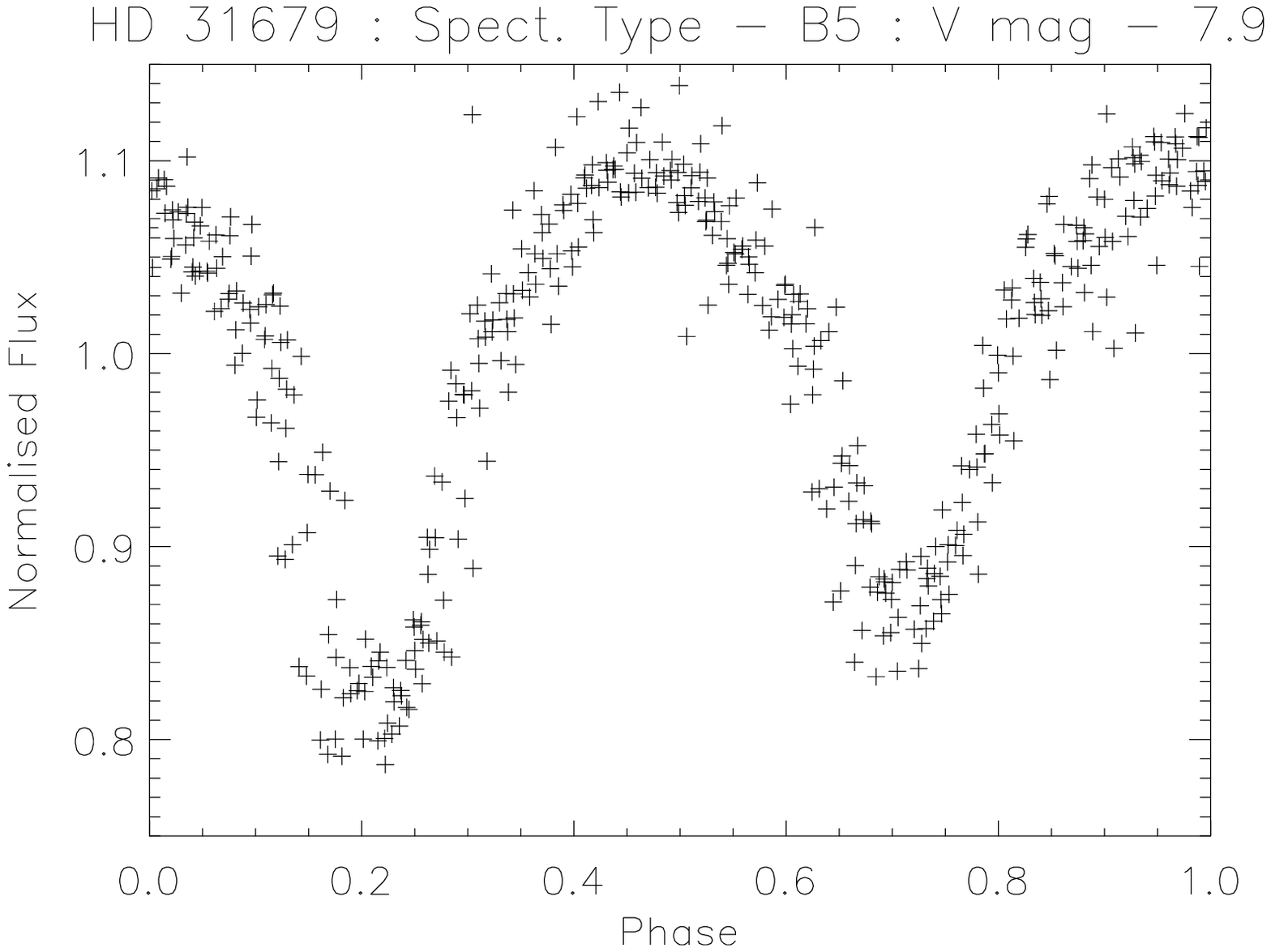}\\
\includegraphics[scale=0.5]{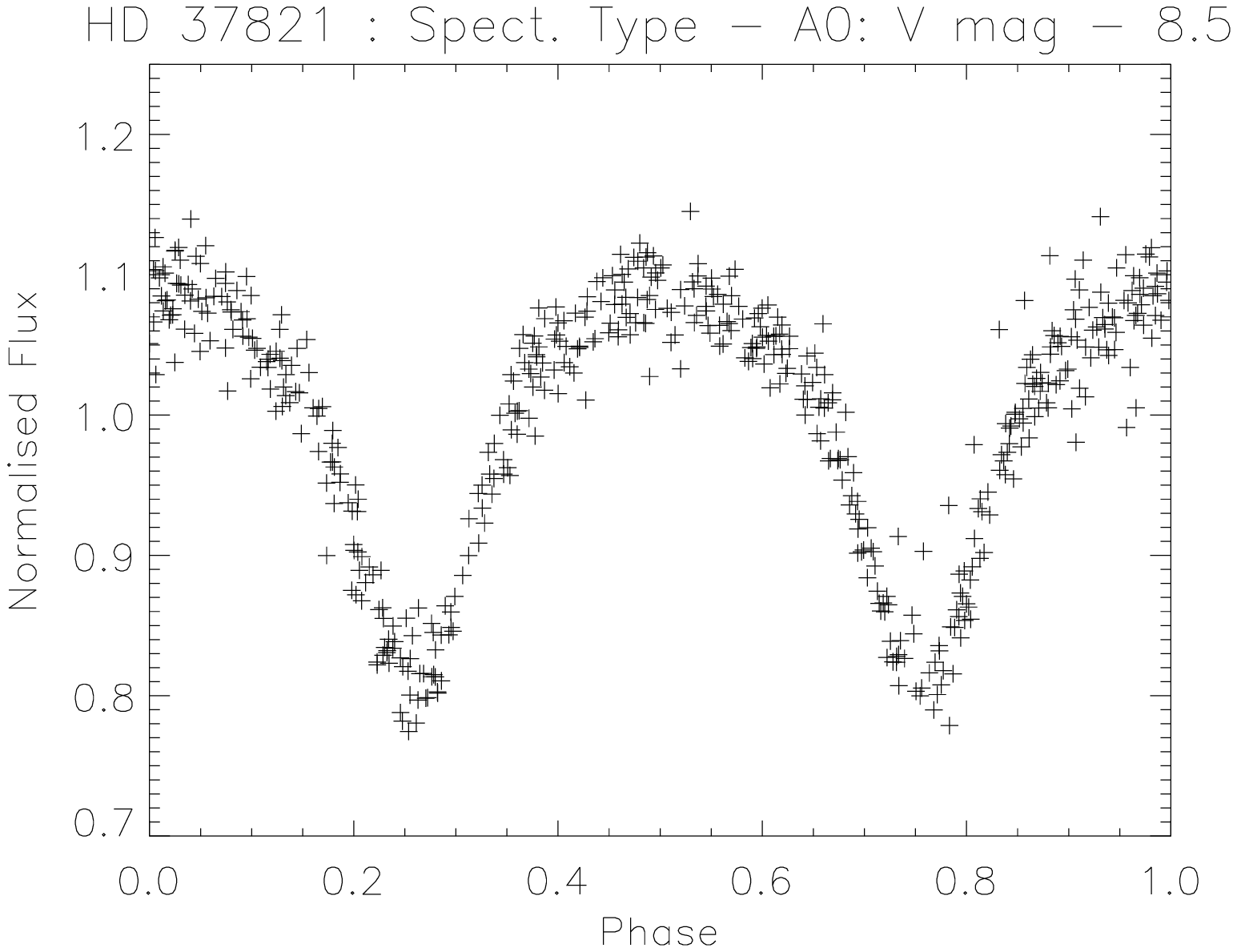}&\includegraphics[scale=0.5]{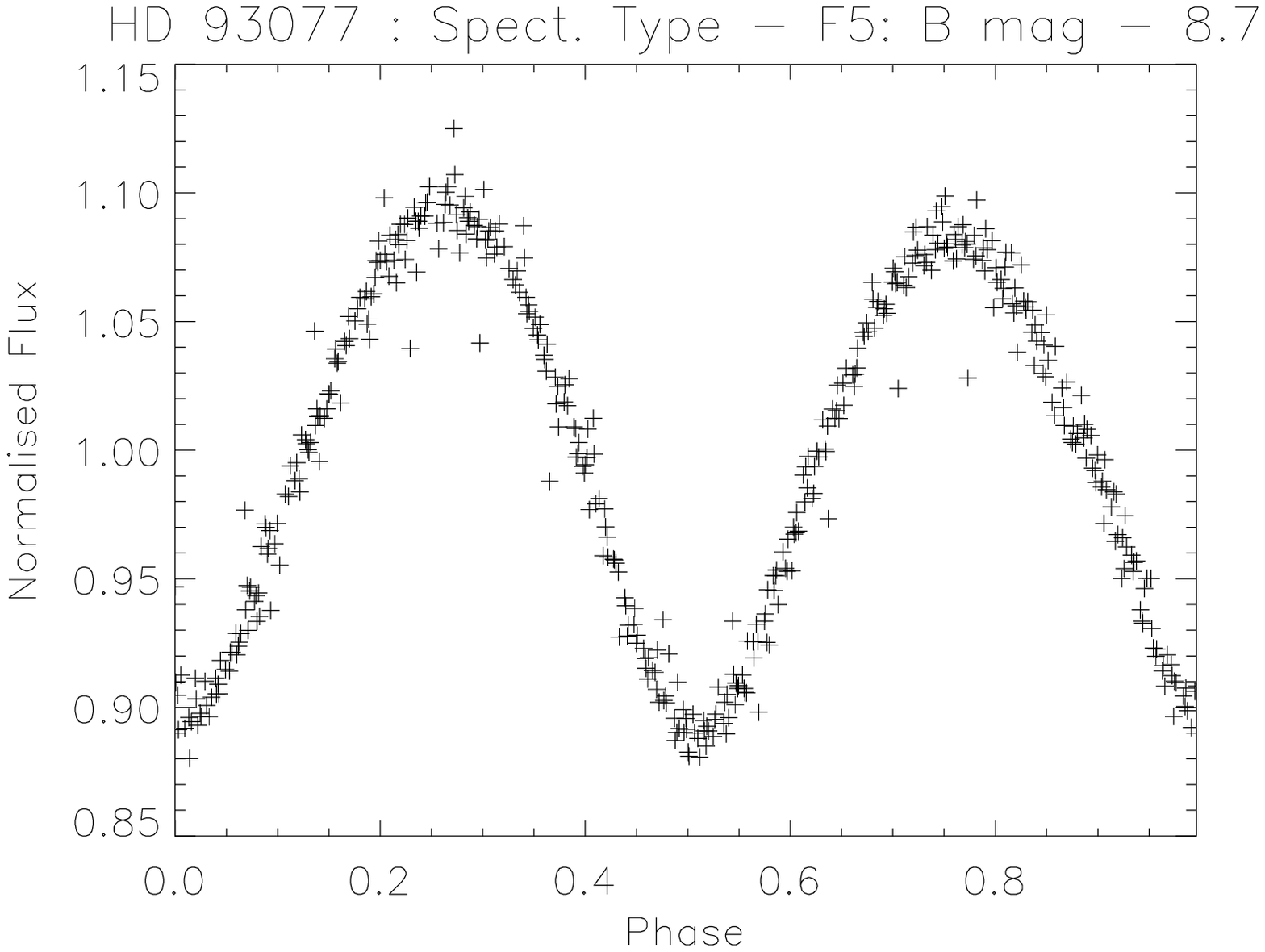}\\
\end{tabular}
\caption{\label{fig:sample_lc} STRESS sample light curves : The above are plots of known variable stars of different types of variability folded on the period found from the STRESS light curve. Again we are stressing that this is only a sample of our data and even these data needs to be further analysed to realise the full potential of the science that can be derived from them.}
\end{figure*}

\section{Results}

\begin{figure*}
\centering
\begin{tabular}{c c}
\includegraphics[scale=0.5]{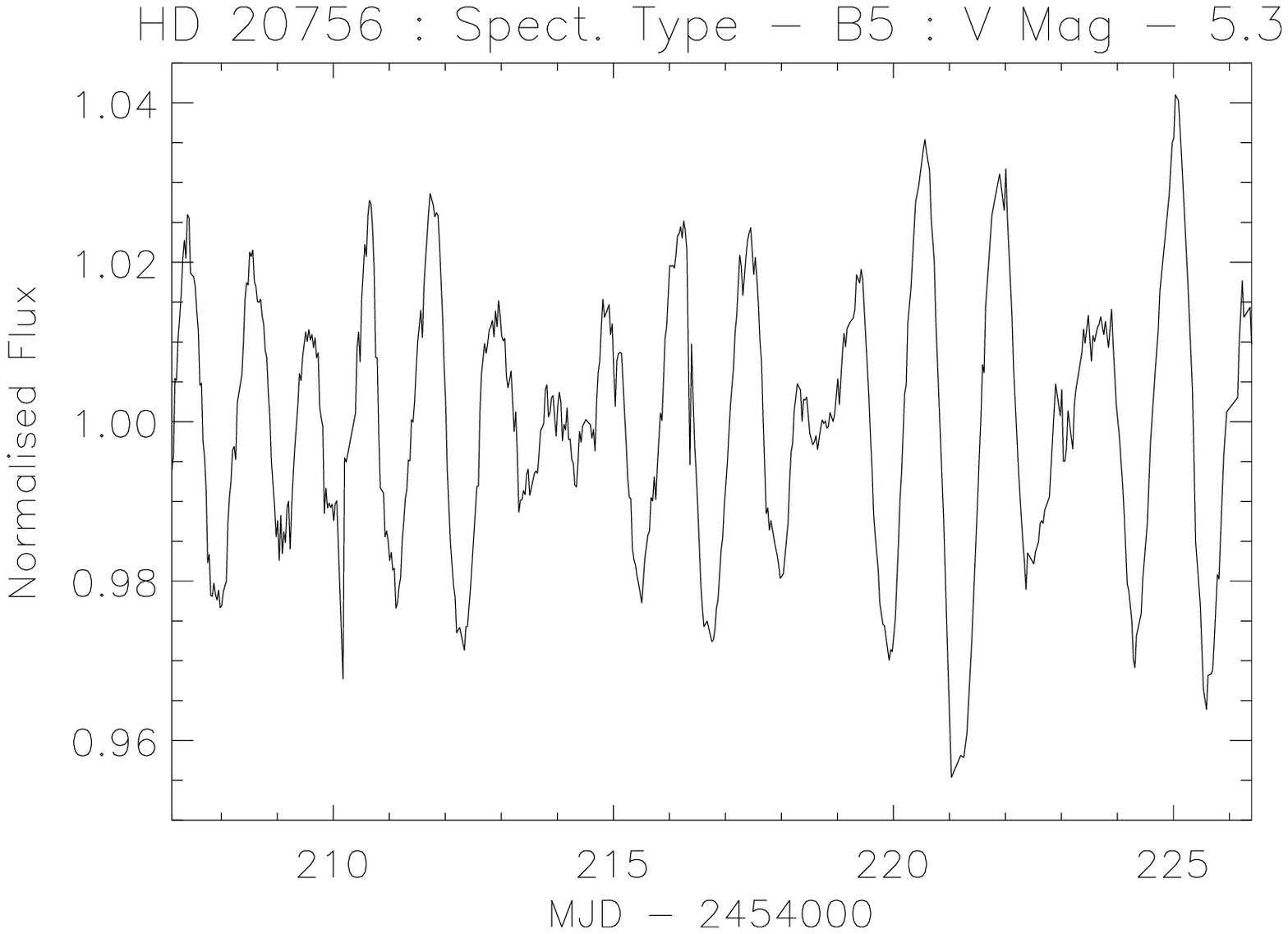}&\includegraphics[scale=0.5]{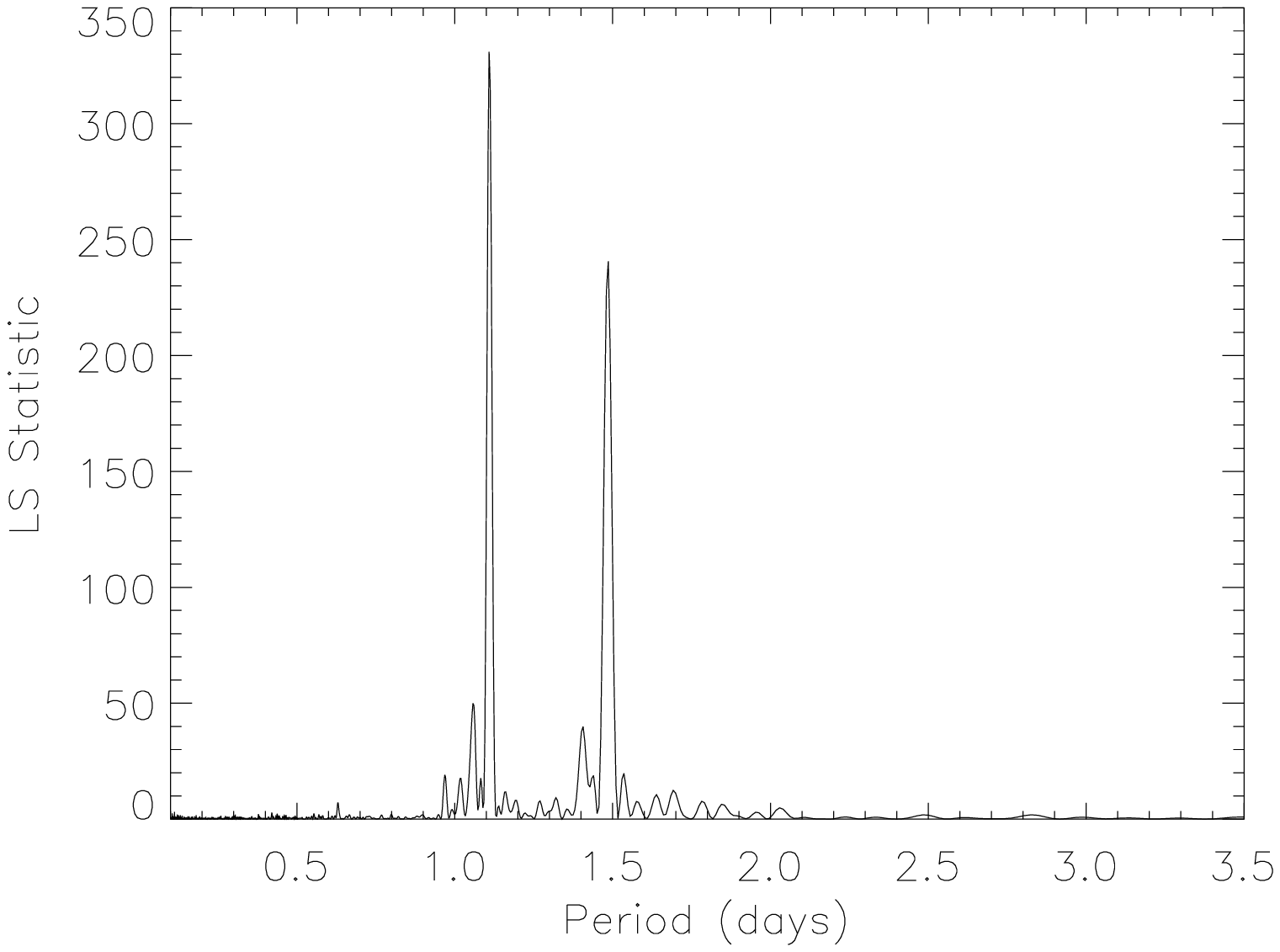}
\end{tabular}
\caption{\label{fig:new_variables}HD 20756 exhibits two prominent period in the STRESS light curve. The left plot is a light curve for one epoch of our data and the right is the LS periodogram. The data exhibits two clear periods at 1.105d and 1.492d.}
\end{figure*}

\indent \indent In this section, we present some of the sample light curves which are obtained by the above analysis pipeline. Fig.~\ref{fig:44_tau},\ref{fig:sample_lc} and \ref{fig:new_variables} \footnote{\bf{Normalised flux in these figures are the actual flux values divided by the median values and added one.}} are a good but random representation of the whole data in terms of magnitude, different variabilities observed as well as quality of the light curve in terms of scatter. {\bf{The left plot in Fig.~\ref{fig:44_tau} shows the whole data set for the three epochs, each made up of 20 days each and the right plot is a blow-up of a few days of data where the variability is clearly evident.}} Further individual study and analysis of the data will be presented in a sequel to this paper.  \\

\indent{\underline{\bf{HD 26322}}}: (= 44 Tau) A F2 star, $\delta$ Scuti variable with 49 known pulsational frequencies (\emph{V} mag = 5.39) in 15 independent modes (\citet{Lenzall} and references therein). The evolutionary phase of this well studied object is still unclear and under investigation. STEREO has data for three epochs (Left plot in Fig.~\ref{fig:44_tau}) and we were able to identify at least 15 pulsational frequencies using Period04 \citep{Lenz}.\\

\indent{\underline{\bf{HD 14147}}}: (= AD Ari) A F0 $\delta$ Scuti variable of \emph{V} mag of 7.43. Even-though \citet{Zhou}, identifies two different periods, \citet{Rodriguez} find only one period at 0.2699d. Our analysis using Period04 gives a value of 0.269d but a phase folded plot at this period, gives two distinct amplitudes. The top right plot of Fig.~\ref{fig:sample_lc} is folded over twice the period found (0.54d) and from this plot, we suspect that this could be a binary system rather a $\delta$ Scuti variable. \\

\indent{\underline{\bf{HD 31679}}}: (= V1061 Tau) A recently discovered \citep{Terrell} B5, eclipsing binary of the $\beta$ Lyrae type with a \emph{V} = 7.95. They found it to be a system with a mass ratio of 2.407 and also speculate on the probable presence of a third light in the light curve, which is not confirmed yet. STRESS light curve exhibits a period of 1.379d.\\

\indent{\underline{\bf{HD 37821}}}: (= V1238 Tau) An A0 eclipsing binary of the W Uma type of a \emph{V} = 8.5. The All Sky Automated Survey Catalogue of Variable Stars (ACVS) defines a period of 1.12173d.\\

\indent{\underline{\bf{HD 93077}}}: (= EX Leo) A F6 eclipsing binary star of the W Uma Type with a known period of 0.40860d with a \emph{B} mag of 8.71. \citet{Eker}, used this star as a sample in their re-calibration of the Period - Luminosity Colour relationship of the W Uma type Binaries. It was also studied by the N2K consortium using the KPNO low resolution spectra in order to identify targets for the Keck Planet search program \citep{Robinson}.\\

\indent\indent Below, we present a sample of new variables (Fig.~\ref{fig:new_variables}) discovered in our data using  our variability analysis pipeline. This analysis algorithm and complete catalogue will be discussed in a forthcoming paper in this series. Hence we would like to stress the fact that these results are presented to just give a flavour to the reader and needs further investigation.\\

\setcounter{figure}{12}
 \begin{figure*}
\begin{tabular}{c c}
\includegraphics[scale=0.5]{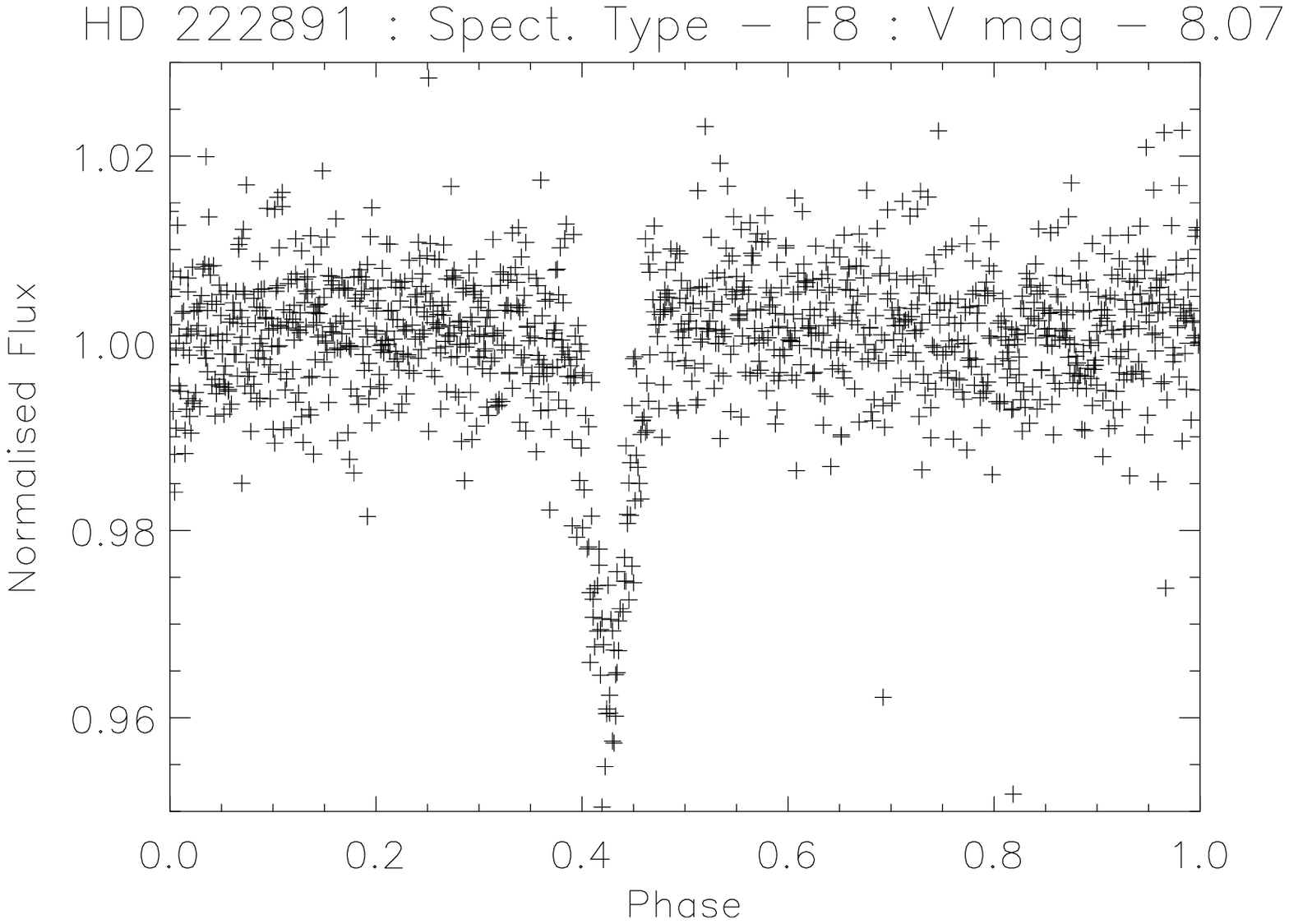}&\includegraphics[scale=0.5]{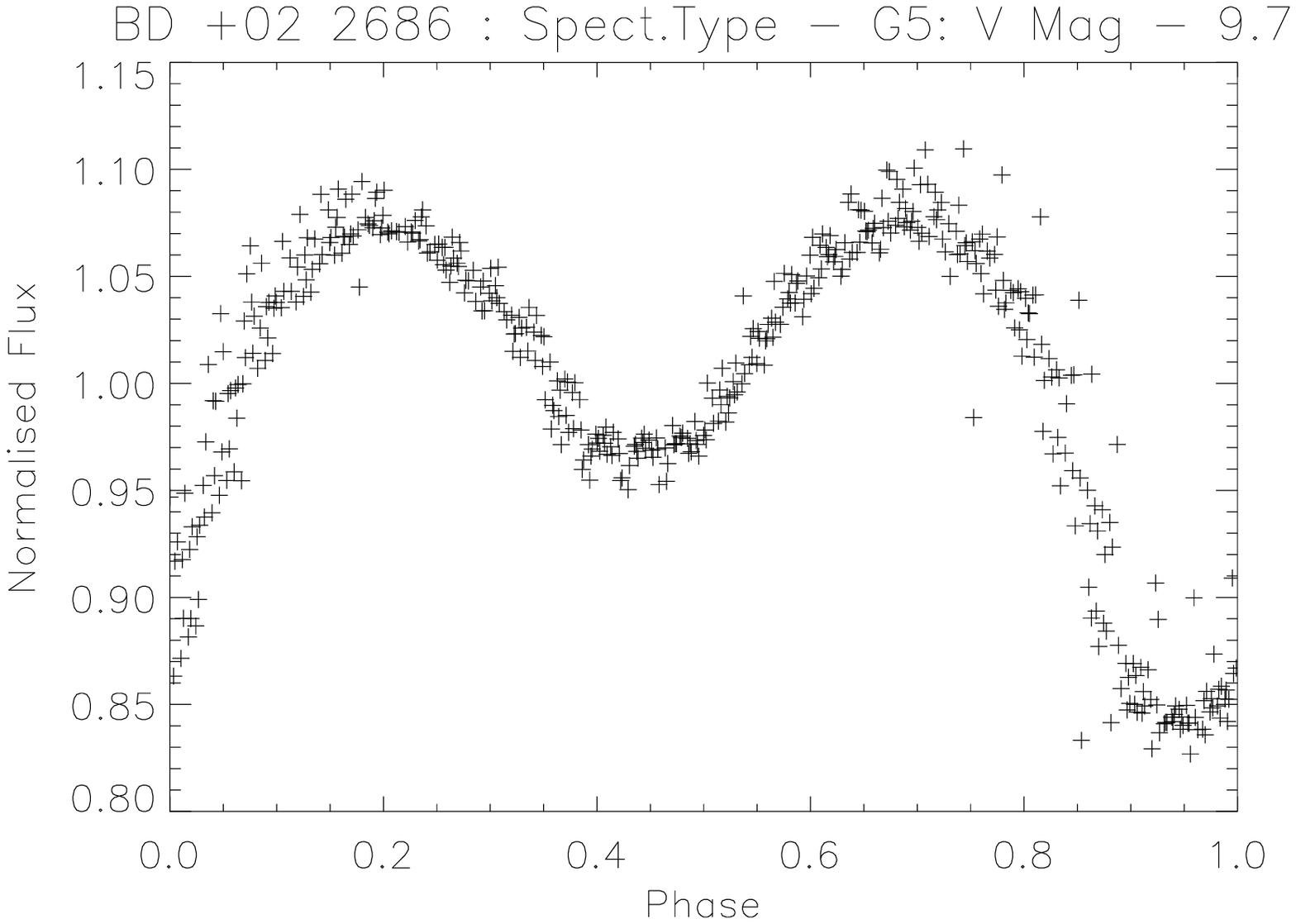}
\end{tabular}
\caption{Two new variables with no previous literature on its variability. The left is a plot of HD 222891 with shallow eclipses of period 1.5949d and also with out-of-eclipse pulsations. The right is a plot of BD+02 2686, a faint eclipsing binary with no known literature, having a period of 0.77d.}
\end{figure*}

\indent{\underline{\bf{HD 20756}}}: (= 61 Tau) A B5 star of \emph{V} mag 5.3 in a multiple system as per Simbad's classification and catalogued as an eclipsing binary by \citet{Eggleton}. From our light curve, we find two prominent periods at 1.105d and 1.492d. As you can see these indicate a controversy with the known literature and needs further investigation.\\

\indent{\underline{\bf{HD 222891}}}: With a \emph{V} =  8.07 and spectral type F8, this star was studied by the N2K consortium as a potential target for the Keck planet search program but was rejected based on its atmospheric parameters \citep{Robinson}. We find well defined eclipses with period of 1.5949d, which are just out of bounds, i.e., too deep for a planetary companion. This light curve also illustrates that we can detect shallow eclipses of less than half percent easily in our data. The presence of additional out-of-eclipse pulsations, nevertheless makes this an interesting system to solve. \\

\indent{\underline{\bf{BD +02 2686}}}: Known as a G5 star of \emph{V} = 9.75 with no other literature. Our light curve clearly shows the presence of both primary and secondary eclipses with a primary period of 0.774d.\\

\section{Summary}

\indent \indent In this paper, we have described the semi-automated stellar photometry pipeline developed for the reduction of \emph{STEREO} HI images. Using this pipeline we have collected photometric information for nearly a million stars covering 37 months from April 2007 till March 2010, with most stars having data for three epochs and some for four epochs. We have also outlined the general characteristics of the data along with some of the systematic effects found in the data and the different methods used to remove these systematic trends. Even though the cause of many of these effects are still not understand completely and needs further analysis, the relative scatter of most data brighter than \emph{R} $=$ 8, is less than one percent and for fainter objects a few percent at best. Finally, we present a sample of the interesting light curves in our data. The data which we have accumulated till date have a lot of potential for further study in terms of variability analysis of known, suspected and new, unknown objects as well as for exoplanetary analysis. We also plan to release this data as a photometric catalogue to the general public in the future, due to its unique characteristics, like the sampling rate (18 c$d^{-1}$ and 6 c$d^{-1}$ for HI-1 and HI-2 respectively), sky coverage (60 percent of sky when all instruments combined), magnitude range (brighter than 12) and also its potential to be combined with other \emph{STEREO} HI instruments (180 days of data in total) and hence can be used for long term study as well. And also round-the-clock observation of the spacecraft's unique field of view in the ecliptic, gives STRESS the unique place in the current day exoplanet and stellar variability search programs.  \\

\section*{Acknowledgements}

\indent We would like to thank Steve Spreckley for his timely advice and help during the development of the reduction pipeline. We would like to thank our referee for his careful reading and valuable comments. We would also like to acknowledge the use of Simbad database for individual study of stars and bibliographic references. The photometric reduction described in this paper were performed using the University of Birmingham's  BlueBEAR HPC service,  which was purchased through HEFCE SRIF-3 funds. See \url{http://www.bear.bham.ac.uk} for more details. \\

\bibliographystyle{mn2e}

{}
\label{lastpage}

\end{document}